\documentclass[12pt]{article}  \usepackage{amsmath}
\usepackage{amssymb} \usepackage{euscript}

\usepackage{epsfig}

\usepackage{cite}

\usepackage{alltt}

\usepackage{epic}

\begin{document}   
\allowdisplaybreaks

\begin{titlepage}

\begin{flushright}
{\bf LC-PHSM-2003-003 \\   CERN-TH/2003-009\\ TTP03-01}
\end{flushright}

\hspace{1 mm}
\vspace{1 mm}
\begin{center}{\bf\Large {\boldmath Measuring the Higgs boson parity 
 }}
\end{center}
\begin{center}{\bf\Large {\boldmath at a Linear Collider }}
\end{center}
\begin{center}{\bf\Large {\boldmath  using the {\boldmath $ \tau$}
impact parameter and {\boldmath $\tau \to \rho \nu$} decay ${\;}$
}}\end{center}
\vspace{0.3 cm}
\begin{center}
  {\large\bf K. Desch$^{a}$,  Z. W\c{a}s$^{b,c}$} ~{\large \bf and}~
  {\large\bf   M. Worek$^{d,e}$ }
\vspace{0.3 cm} \\{\em $^a$ Universit\"{a}t Hamburg, Institut f\"{u}r
Experimentalphysik\\  Luruper Chaussee, D-22761 Hamburg, Germany. }\\
{\em $^b$ The Henryk Niewodnicza\'nski Institute of Nuclear Physics\\
Radzikowskiego 152, 31-342 Cracow, Poland.}\\   {\em $^c$ CERN, Theory
Division, CH-1211 Geneva 23, Switzerland.}\\  {\em $^d$
Universit\"{a}t Karlsruhe, Institut f\"{u}r Theoretische
Teilchenphysik   \\ D-76128 Karlsruhe, Germany.}\\ {\em $^e$ Institute
of Physics, University of Silesia\\ Uniwersytecka 4,    40-007
Katowice, Poland.}\\
\end{center}
\vspace{1mm}
\begin{abstract}
We demonstrate that a measurement of the impact parameter in one-prong
$\tau$ decay can be  useful for the determination of the  Higgs boson
parity in the $H/A\to\tau^{+}\tau^{-}$;
$\tau^{\pm}\to\rho^{\pm}\bar{\nu}_{\tau}(\nu_{\tau})$ decay chain.  We
have estimated that for a detection set-up such as TESLA, use of the
information from the $\tau$ impact parameter can improve the significance
of the measurement of the parity of the Standard Model $120$ GeV Higgs
boson to $\sim$ 4.5$\sigma$, and in general  by  factor of about
1.5 with respect to the method where this information is
not used.

We also show that the variation in the assumption on the precision of the
measurement of the impact parameter and/or $\pi$'s momenta does not affect
the sensitivity of the method. This is because the  method  remains
limited by the type of  twofold ambiguity in reconstructing the
$\tau$ momentum.

\end{abstract}
\begin{center}
{\it Submitted to Eur. Phys. J.  }
\end{center}

\vspace{3mm}
\vfill
\begin{flushleft}
{\bf 
January 2003}
\end{flushleft}

\end{titlepage}
\section{Introduction}

One of the most important goals in the scientific program of a future
linear electron--positron collider (LC)   is to measure the
properties of the Higgs boson  precisely.  Among them, the parity plays a
prominent role.  Depending on the mass and the Higgs  mechanism, the
Higgs parity can be measured by different  means
\cite{Barger1994,Hagiwara1994,Skjold:1994qn,Boe:1998kp,Hagiwara2000}. In
some cases, it can be best determined  by investigating the properties
of the Higgs boson decay into $\tau$  leptons
\cite{Kramer:1994jn,Grzadkowski:1995rx}.   In \cite{Bower:2002zx} we
studied the possibility of distinguishing a scalar from a pseudoscalar
coupling of a light boson  to fermions using its decay to a pair of
$\tau$ leptons and their  subsequent decays
$\tau^{\pm}\to\rho^{\pm}\bar{\nu}_{\tau}(\nu_{\tau})$ and
$\rho^{\pm}\to\pi^{\pm}\pi^{0}$.  This decay  chain gives an independent test of model and
production mechanism.  In fact, we have found that
the angular distributions of the  $\tau^{\pm}$ decay products, which
clearly distinguish the different  parity states, are measurable using
typical properties of a future  detector at an $e^{+}e^{-}$ linear
collider.   A special Monte Carlo technique was developed  for that
purpose \cite{Was:2002gv}.  In the present paper we will extend that
analysis, using the information that can be obtained  from the
measurement of the $\tau$ lepton impact parameter.

  It turns out that the kinematic configuration of
$e^{+}e^{-}\to\tau^{+}\tau^{-}$ events cannot be fully determined
since the momenta of both $\nu_{\tau}$ and $\bar{\nu}_{\tau}$ are
unknown.  The lost information can be recovered  partially by
measuring  energy and   directions of all visible particles
\cite{Kuhn:1982di,Hagiwara:1989fn}.   When both $\tau$ leptons decay
hadronically and all hadron momenta are recovered,  the original
$\tau$ direction can be obtained up to a twofold  ambiguity
\cite{Kuhn:1982di,Tsai:1965}.  In the method, the requirement that the
two $\tau$ leptons be essentially  of equal energy and back to back
in some known frame,  is decisive.  The measurement of the tracks of
the hadrons, in particular their relative impact parameters, allows us to
resolve this twofold ambiguity, see for instance
\cite{Kuhn:1982di,Hagiwara:1989fn,Kuhn:1993ra}.

In the case of linear colliders such as TESLA
\cite{Aguilar-Saavedra:2001rg}, a sufficiently precise reconstruction of
the frame,  where $\tau$ leptons are expected to be back to back turns
out to be impossible,  mainly due to beamstrahlung effect. This is
a strong limitation  for the Higgs boson parity measurement as proposed
in Refs.   \cite{Kramer:1994jn,Grzadkowski:1995rx}. A way around was
found in Ref.  \cite{Bower:2002zx}.  Here, we will investigate if  the
measurement of the $\tau$ lepton impact parameter direction, with
precision as expected for TESLA, can be used to optimize that method.

This work is organized as follows: in section 2 the geometry of $\tau$
production and decay  is described, as well as our way to use the $\tau$
impact parameter in the measurement of  the Higgs boson parity.  In
section 3  some details of the {\tt TAUOLA} Monte Carlo simulation are
given. Section 4 is devoted to the definitions  used in our simulation
of  detector properties. In section 5 definitions of our observable and
numerical results are given. The summary, section 6, closes the paper.

\section{Description of the method}
\begin{figure}[!ht]
\begin{center}  
\epsfig{file=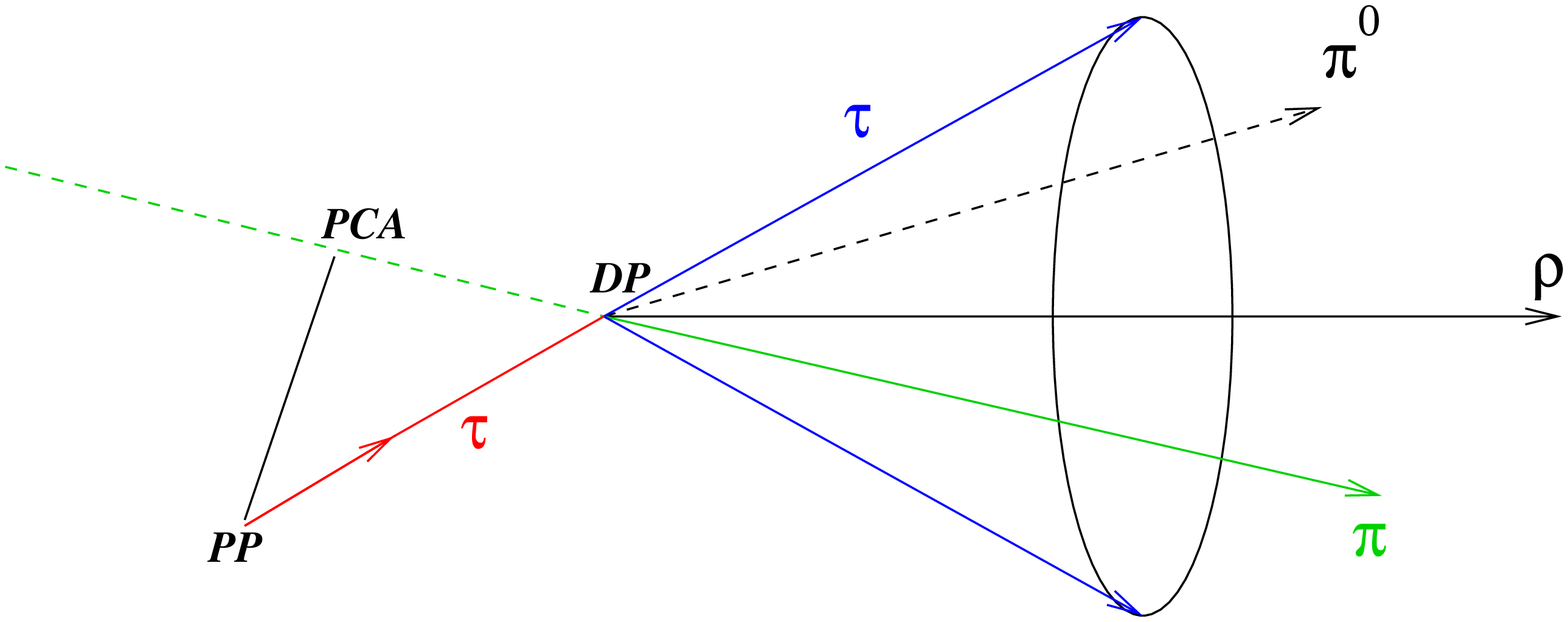,width=140mm,height=40mm}
\end{center} 
\caption  
{\it Schematic view  of the $\tau$ decay to $\rho$ and $\nu_{\tau}$.
$PP$ is the $\tau$ production point, $DP$ the $\tau$ decay point, and $PCA$
the point of closest approach. Directions and energies of $\pi^{\pm}$,
$\pi^0$, $\rho^{\pm}$ are directly measurable, the  energy of the $\tau^\pm$
can be reconstructed from the event topology. The ${\tau}$ momentum
must be placed on the cone (drawn on the plot) and the $\tau$ impact
parameter can be used to determine the actual position.  }
\label{figure1}  
\end{figure}

In the decay of the $\tau$ lepton into one charged particle, the
$\tau$ impact parameter can be defined as the distance of closest
approach (in  the plane perpendicular to the beam axis) of the charged
particle to a reference  point, which is assumed to be the production
point of the $\tau$ (see Fig.~\ref{figure1}). A positive sign is given
to this quantity if the track crosses the $\tau$ direction in the same
hemisphere, relative to the $\tau$ production point,  in which it lays,
and negative otherwise. The production point is assumed  to lie on the
beam line. In our study we will exploit the possibility to determine the
direction of the impact parameter, rather than the related distance.
In fact we will be interested, where the $\tau$ production point is
localized with respect to the charged pion track.

Before we  address the question of the information available from
the  impact parameter, let us recall some details of the actual
observable  proposed  in Ref.\cite{Bower:2002zx} for the Higgs boson parity
measurement. The method relies on measuring the acoplanarity angle of the
two planes, spanned on  $\rho^{\pm}$ decay products and defined  in
the $\rho^{+}\rho^{-}$ pair rest frame; some additional selection cuts also
need to be applied.  Let us present the elements of the observable 
as the following points.

\vspace{1 cm}
\vspace{1 cm}
\begin{figure}[!ht]
\begin{center}  
\epsfig{file=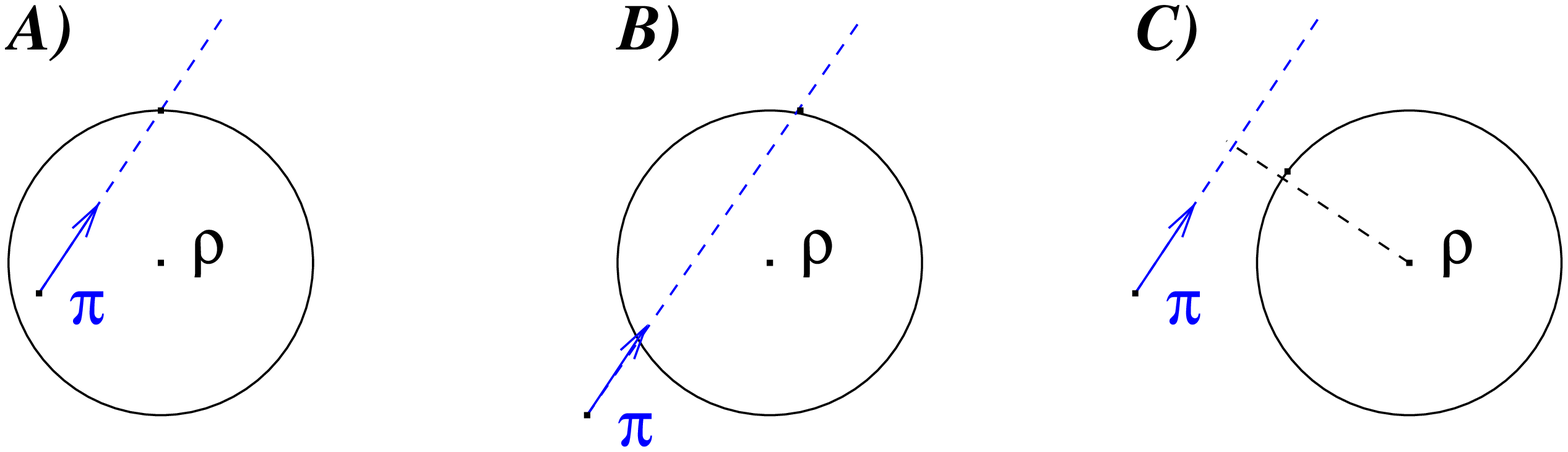,width=140mm,height=40mm}
\end{center}
\caption  
{\it  The circle of candidates for $\tau$ momentum.  Depending on the
 possible geometrical configuration, three cases are possible:  A) $\pi$
 direction inside the circle: single candidate for $\tau$ momentum.
 B) $\pi$ direction outside the circle: two candidates for $\tau$
 momentum. C)  $\pi$ direction outside the circle and no crossing of
 dashed line with the circle (due to detection ambiguities of measured
 angles and energies). We take in this case the closest point  to the
 line. Effectively, we obtain a single solution.   }
\label{figure2}  
\end{figure}  
\vspace{-1 cm}
\begin{enumerate}
\item{ \bf Acoplanarity of the {\boldmath $\rho^{+}$} and  {\boldmath
$\rho^{-}$} decay planes:}
\begin{itemize}
\item The reconstructed four-momenta of $\pi^{+}$ and $\pi^{0}$ are
combined to yield the $\rho^{+}$ four-momentum. The same is done for the
$\rho^{-}$.
\item All reconstructed four-momenta are boosted into the
$\rho^{+}\rho^{-}$ rest frame.
\item The angle $\varphi$ between the planes of the $\rho^{+}$ and
$\rho^{-}$ decay products in this frame is the acoplanarity.
\end{itemize}
\item{\bf Normalized energy differences:}\\ The events are divided
into two classes depending on the value of  $y_{1}y_{2}$, where
\begin{equation}
\label{selection}
y_1={E_{\pi^{+}}-E_{\pi^{0}}\over E_{\pi^{+}}+E_{\pi^{0}}}~;~~~~~
y_2={E_{\pi^{-}}-E_{\pi^{0}}\over E_{\pi^{-}}+E_{\pi^{0}}}.
\end{equation}
The energies of $\pi^\pm,\pi^0$ are to be taken in the  respective
$\tau^\pm$ rest frames.
\item{\bf Replacement {\boldmath $\tau$} rest frame:}\\ In
Ref.~\cite{Bower:2002zx} the method of reconstruction of {\it replacement
$\tau^\pm$ rest frames} was proposed. We will use this method here as
well, but we will omit the  details. In fact they are rather
unimportant. The energies of pions taken directly  from the laboratory
frame can  equally well be used in the above  definitions of Eq.~(\ref{selection}).
\item{\bf Higgs rest frame:}\\ If the information on the beam energies
and the energies of all other observed particles is taken into
consideration, the Higgs rest frame  can be reconstructed,
for instance in the Higgsstrahlung production process, $e^+e^-\to H Z$,
when the $Z$ boson decays either into a charged lepton pair or
hadronically.
We may  define the `reconstructed' Higgs boson  momentum as the
difference of the sum of beam momenta and momenta of all visible
particles, { i.e.} decay  products of $Z$ and all radiative photons
of $| \cos\theta| < 0.98$.
\item{\bf {\boldmath $\tau$} energy:}
\begin{itemize}
\item  In the reconstructed Higgs boson rest frame, the $\tau$
four-momenta are estimated in a crude way by assuming the direction of
the respective  $\rho$'s and an energy of $m_{H}/2$.
\item The $\tau$ momenta are boosted back to the laboratory frame and
their  energies are taken to be the reconstructed energies of $\tau$
leptons.
\end{itemize}
\item{\bf {\boldmath $\tau$} direction:}\\ The $\tau$ direction in the
laboratory system is constrained by two requirements:
\begin{itemize}
\item It has to lie on a cone around the $\rho$ direction with opening
angle $\psi$ defined as:
\begin{equation}
\label{equation}
\cos\psi=\frac{2E_{\tau}E_{\rho}-m^{2}_{\tau}-m^{2}_{\rho}}{2\vec{p}_{\tau}\vec{p}_{\rho}},
\end{equation}
where $E$, $\vec{p}$, $m$ are the energy, momentum and mass,
respectively.  The  four-momentum of $\rho$ is  measured as the sum of
the four-momenta of  its charged and neutral daughters.
\item It has to lie in the plane spanned by the vector pointing  from
$PP$ to $PCA$ and the $\pi^{+}$ momentum.
\end{itemize}
The intersection of the cone and the plane is calculated
numerically. There  are three cases:
\begin{itemize}
\item One solution: the solution is taken as the $\tau$ direction (see
Fig.~\ref{figure2} case A).
\item Two solutions: one of  the two solutions (see Fig.~\ref{figure2}
case B)   is taken on a random basis.
\item No solution: the direction on the cone closest to the
$PCA$--$\pi^{+}$ momentum plane is taken (see Fig.~\ref{figure2} case C).
\end{itemize}
\item{\bf  Impact parameter improved {\it replacement { \boldmath
$\tau$}-rest frame}:}\\  With  the help of the   $\tau$ energy and $\tau$ momentum
defined in points 5 and 6 above, a new impact parameter
improved {\it replacement $\tau$ rest frame} can be defined.   In this way
we have an alternative  way of estimating the difference of $\pi^\pm$,
$\pi^0$ energies see Eq.~(\ref{selection}) in $\tau^\pm$ rest frames.

\end{enumerate}

\section{Monte Carlo set-up}

In the following discussion all the Monte Carlo samples have been
 generated with the {\tt TAUOLA} library
 \cite{Jadach:1990mz,Jezabek:1991qp,Jadach:1993hs}.   For the
 production of  $\tau$ lepton pairs, the Monte Carlo program {\tt
 PYTHIA 6.1} \cite{Pythia} was used\footnote{ It was shown that the
 interface can work in  the same manner with the {\tt HERWIG}
 \cite{HERWIG} generator.}. The production process $e^{+}e^{-}\to ZH
 \to \mu^{+}\mu^{-} (q \bar q) H$ has been chosen, with a Higgs boson
 mass of $120$ GeV and a centre-of-mass energy of  $350$ GeV.

 The effects of initial-state bremsstrahlung were included in the {\tt
 PYTHIA} generation.   For  the $\tau^\pm$ lepton pair decay (with
 full spin effects from the $H \to \tau^+\tau^-$, $\tau^{\pm} \to
 \rho^{\pm}\bar{\nu}_{\tau}(\nu_{\tau})$, \
 $\rho^{\pm}\to\pi^{\pm}\pi^{0}$ chain) the interface explained in
 Ref.~\cite{Was:2002gv} was used.  It is an extended version of the
 interface of Refs.~\cite{Pierzchala:2001gc,Golonka:2002iu}.

\begin{figure}[!ht]  
\setlength{\unitlength}{0.1mm}
\begin{picture}(1600,800)  
\put( 375,750){\makebox(0,0)[b]{\large }}
\put(1225,750){\makebox(0,0)[b]{\large }}   \put(-260,
-400){\makebox(0,0)[lb]{\epsfig{file=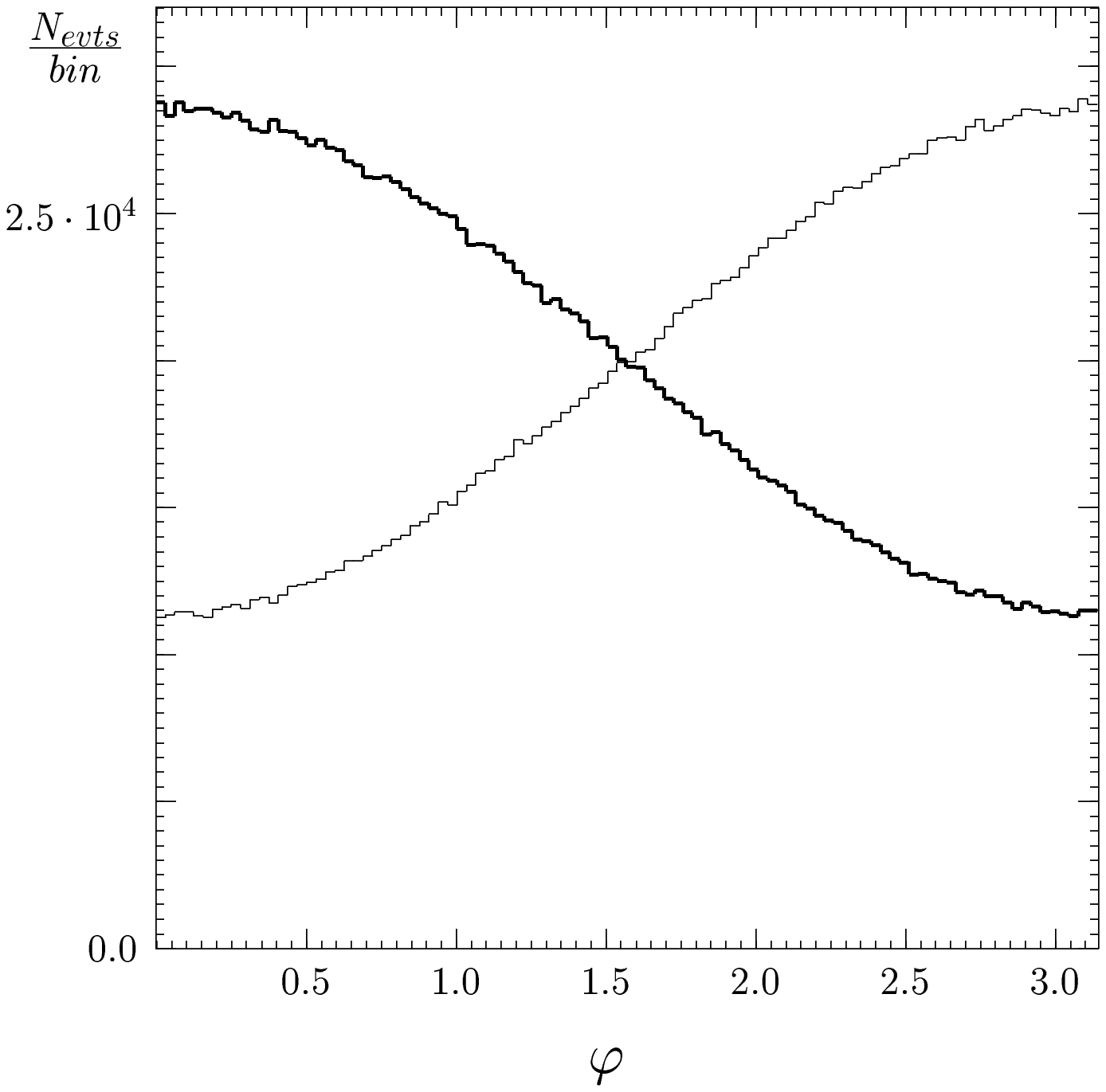,width=120mm,height=140mm}}}
\put(580,
-400){\makebox(0,0)[lb]{\epsfig{file=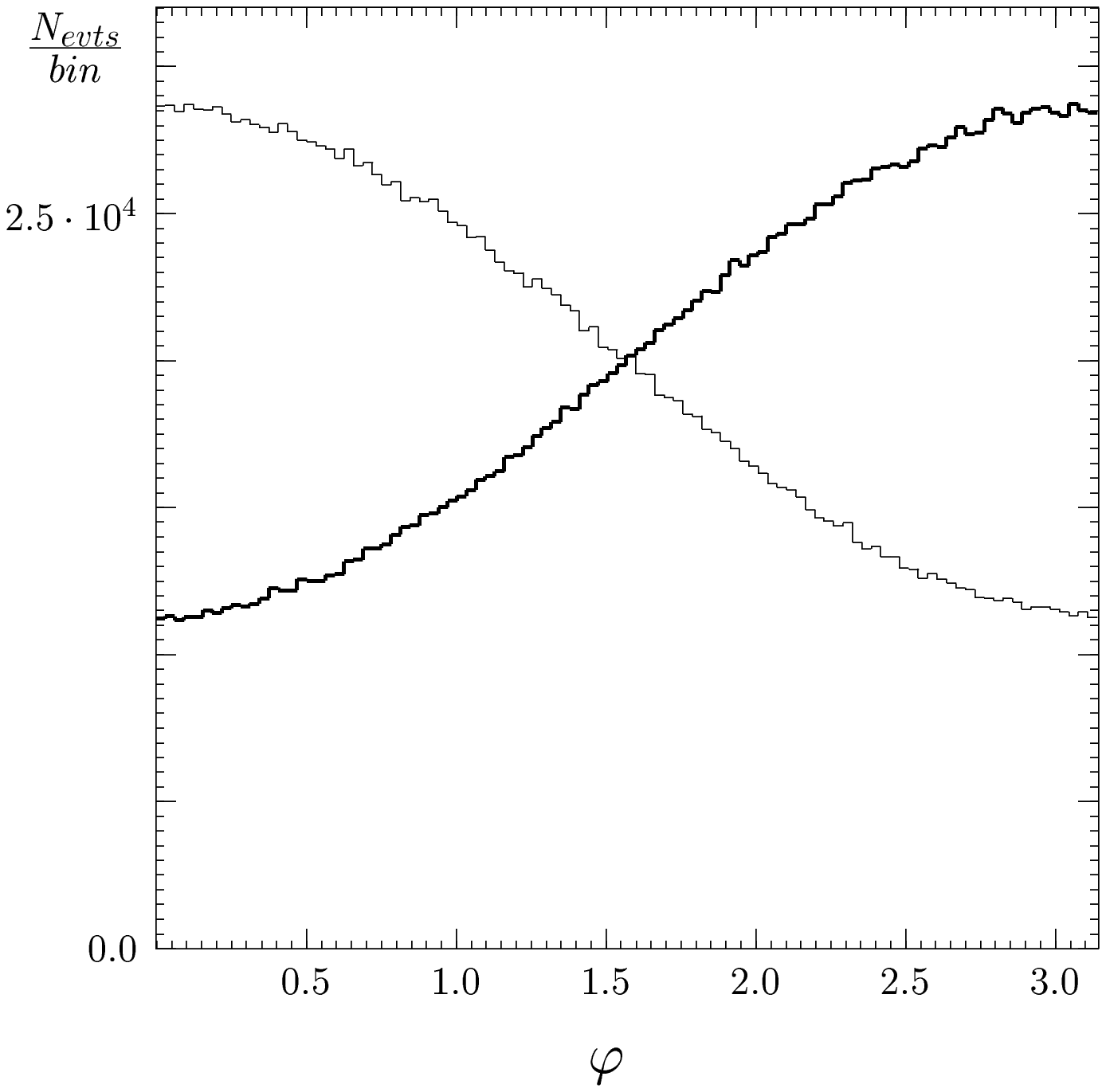,width=120mm,height=140mm}}}
\end{picture}  
\caption  
{\it The acoplanarity distribution
angle $\varphi$ of the $\rho^+ \rho^-$ decay products    
in the rest frame of the $\rho^+ \rho^-$ pair.
Detector smearing is included.  Generator level $\tau^\pm$ rest frames
are used.  The thick line corresponds to a scalar Higgs boson, 
the thin line to a pseudoscalar one.   The left figure contains events with
$y_1 y_2 > 0$, the right is for $y_1 y_2 < 0$. }
\label{aco}  
\end{figure}  
\begin{figure}[!ht]  
\setlength{\unitlength}{0.1mm}
\begin{picture}(1600,800)  
\put( 375,750){\makebox(0,0)[b]{\large }}
\put(1225,750){\makebox(0,0)[b]{\large }}   \put(-260,
-400){\makebox(0,0)[lb]{\epsfig{file=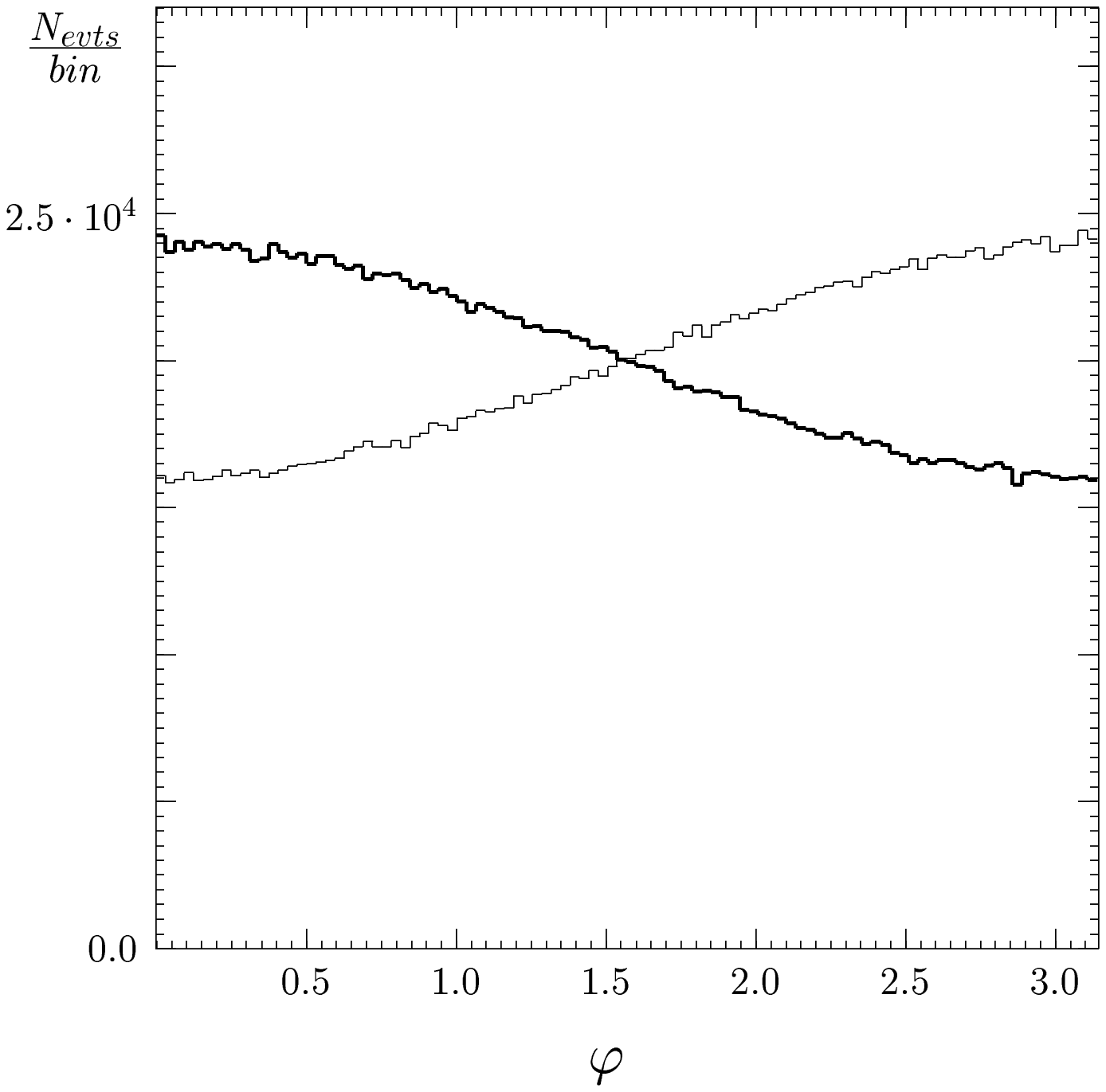,width=120mm,height=140mm}}}
\put(580,
-400){\makebox(0,0)[lb]{\epsfig{file=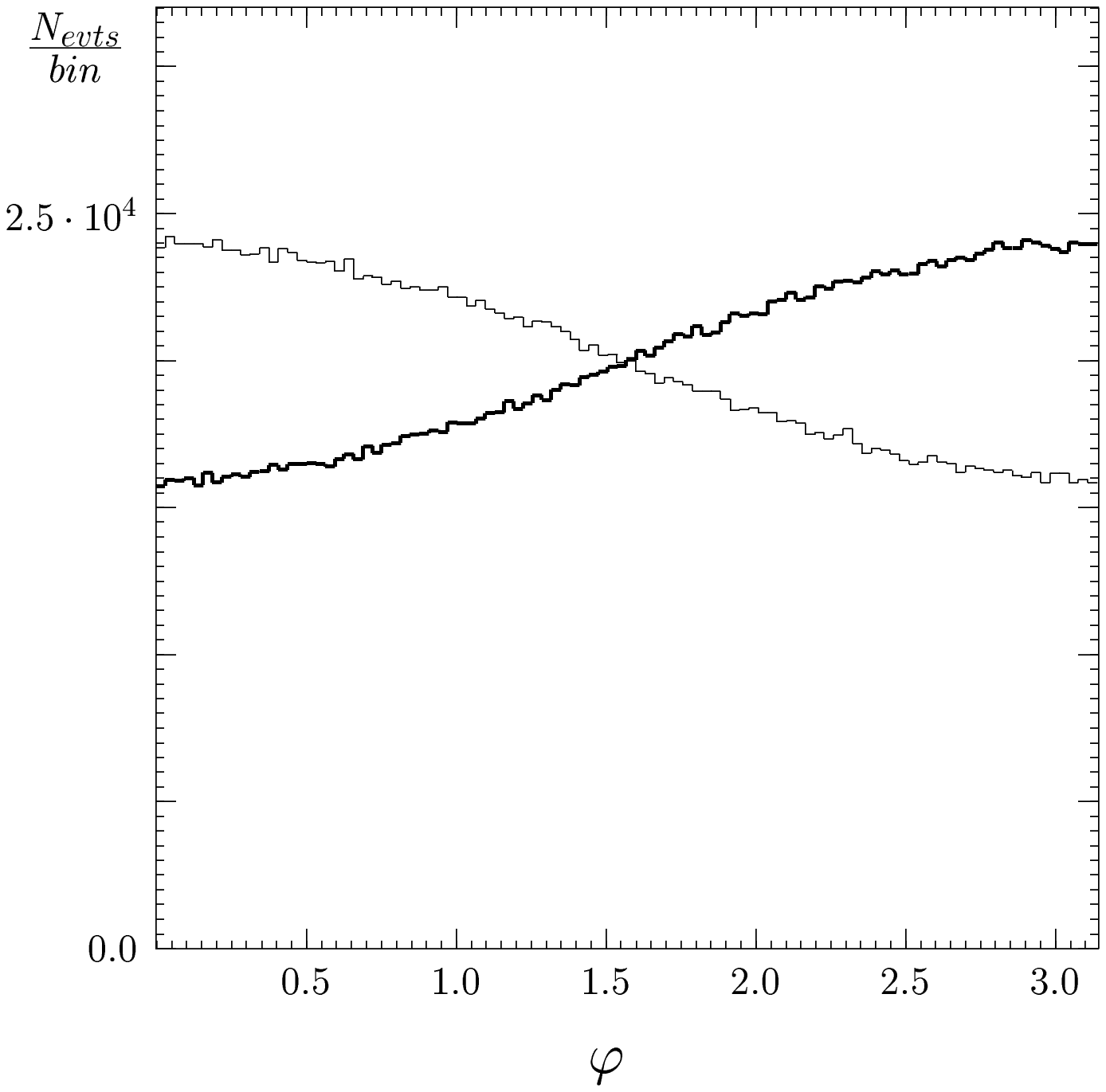,width=120mm,height=140mm}}}
\end{picture}  
\caption  
{\it The acoplanarity distribution
angle $\varphi$ of the $\rho^+ \rho^-$ decay products    
in the rest frame of the $\rho^+ \rho^-$ pair.
Full smearing is included.  Replacement $\tau^\pm$ rest frames defined
as in Ref.~\cite{Bower:2002zx}  are used. The thick line corresponds
to a scalar Higgs boson, the thin line to a pseudoscalar one.
The left figure contains events with $y_1 y_2 > 0$, the right is for
$y_1 y_2 < 0$.
 }
\label{acosm2}  
\end{figure}  
\begin{figure}[!ht]  
\setlength{\unitlength}{0.1mm}
\begin{picture}(1600,800)  
\put( 375,750){\makebox(0,0)[b]{\large }}
\put(1225,750){\makebox(0,0)[b]{\large }}   \put(-260,
-400){\makebox(0,0)[lb]{\epsfig{file=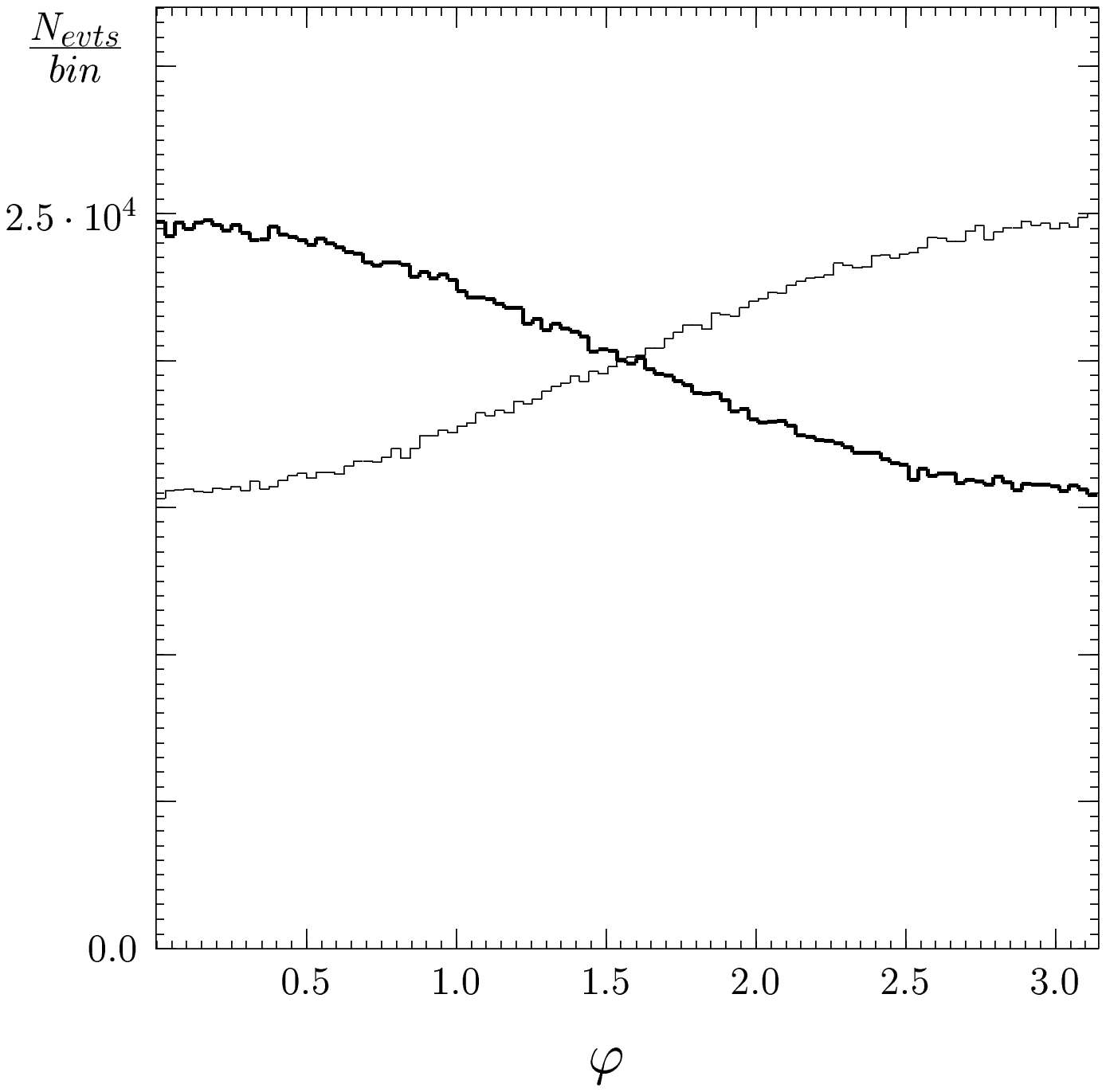,width=120mm,height=140mm}}}
\put(580,
-400){\makebox(0,0)[lb]{\epsfig{file=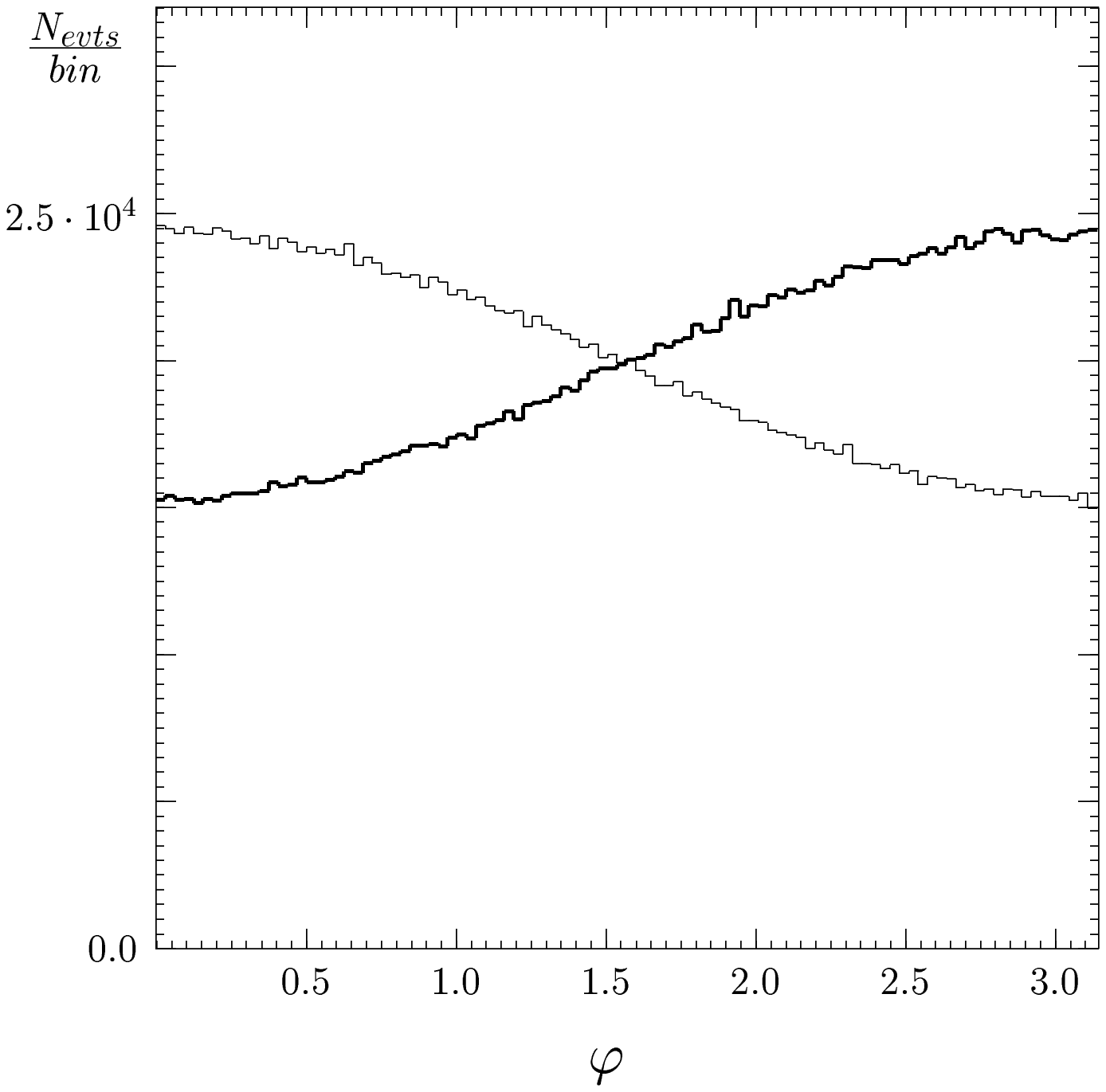,width=120mm,height=140mm}}}
\end{picture}  
\caption  
{\it The $\rho^+ \rho^-$ decay products' acoplanarity distribution
angle, $\varphi$,   in the rest frame of the $\rho^+ \rho^-$ pair.
Full smearing is included.  The $\tau$ lepton impact parameter is used
in the reconstruction of the $\tau^\pm$ rest frames.  The thick line
corresponds to a scalar Higgs boson, the thin line to a
pseudoscalar one.   The left figure contains events with $y_1 y_2 >
0$, the right is for $y_1 y_2 < 0$.}
\label{impact1}  
\end{figure}  

\begin{figure}[!ht]  
\setlength{\unitlength}{0.1mm}
\begin{picture}(1600,800)  
\put( 375,750){\makebox(0,0)[b]{\large }}
\put(1225,750){\makebox(0,0)[b]{\large }}   \put(-260,
-400){\makebox(0,0)[lb]{\epsfig{file=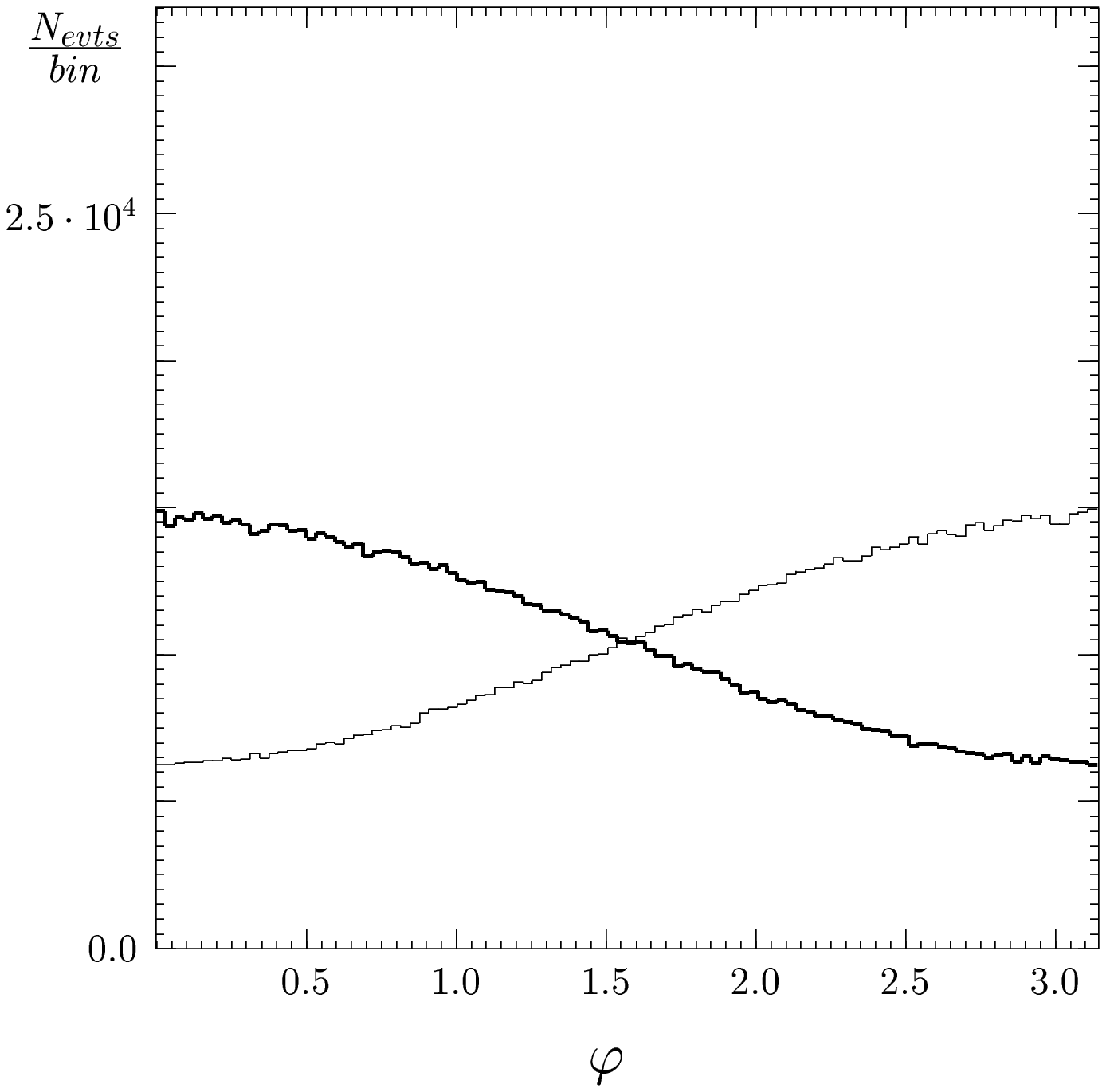,width=120mm,height=140mm}}}
\put(580,
-400){\makebox(0,0)[lb]{\epsfig{file=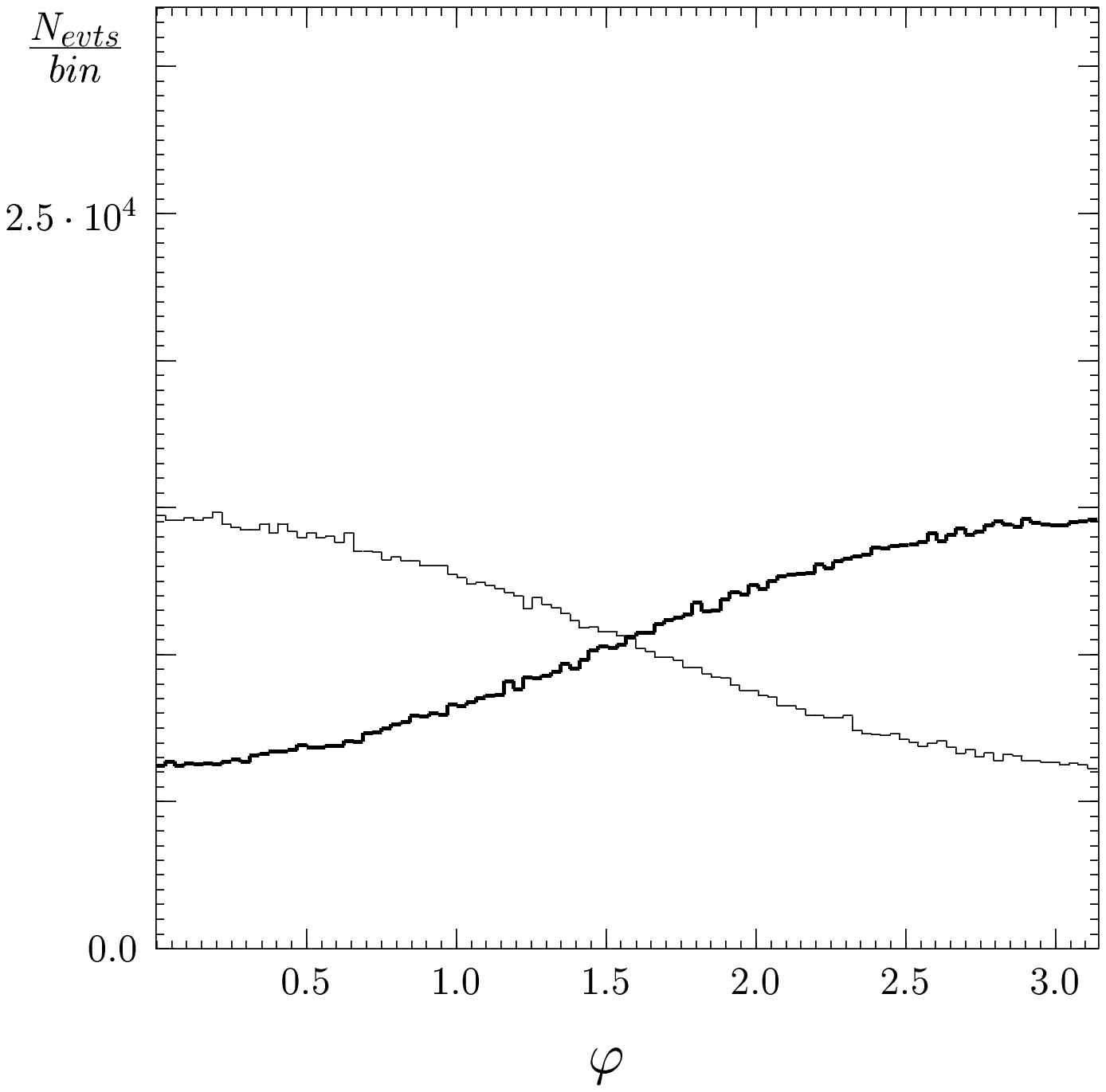,width=120mm,height=140mm}}}
\end{picture}  
\caption  
{\it 
The $\rho^+ \rho^-$ decay products' acoplanarity distribution
angle, $\varphi$,   in the rest frame of the $\rho^+ \rho^-$ pair.
 Only events where the signs of the  energy differences y1 and y2
are the same whether calculated using the method described in
Ref.~\cite{Bower:2002zx} or with the help of the $\tau$ impact
parameter are taken. The thick line corresponds to a scalar Higgs
boson,  the thin line to a pseudoscalar one.    }
\label{impact2}  
\end{figure}  

\begin{figure}[!ht]  
\setlength{\unitlength}{0.1mm}
\begin{picture}(1600,800)  
\put( 375,750){\makebox(0,0)[b]{\large }}
\put(1225,750){\makebox(0,0)[b]{\large }}   \put(-260,
-400){\makebox(0,0)[lb]{\epsfig{file=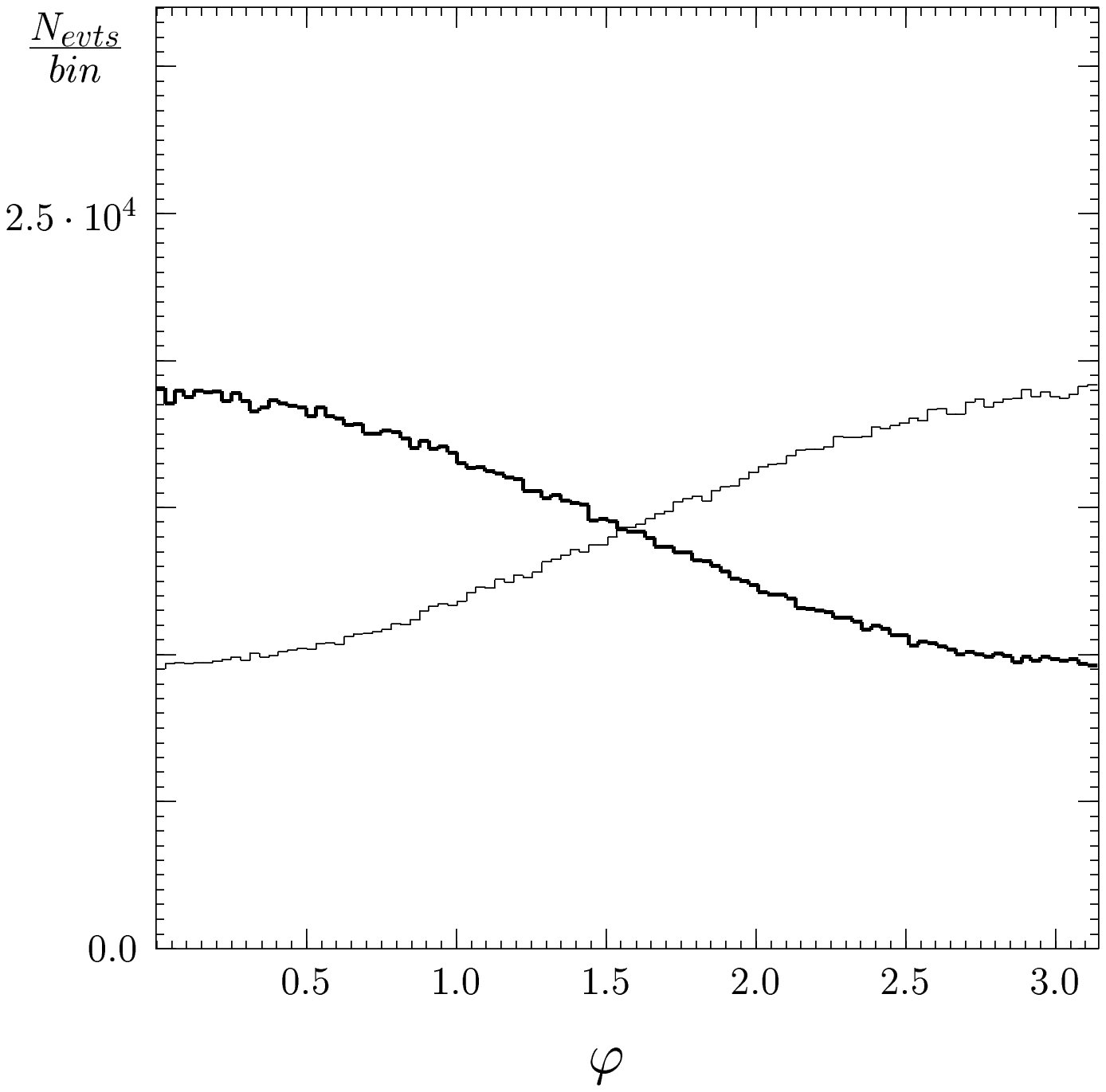,width=120mm,height=140mm}}}
\put(580,
-400){\makebox(0,0)[lb]{\epsfig{file=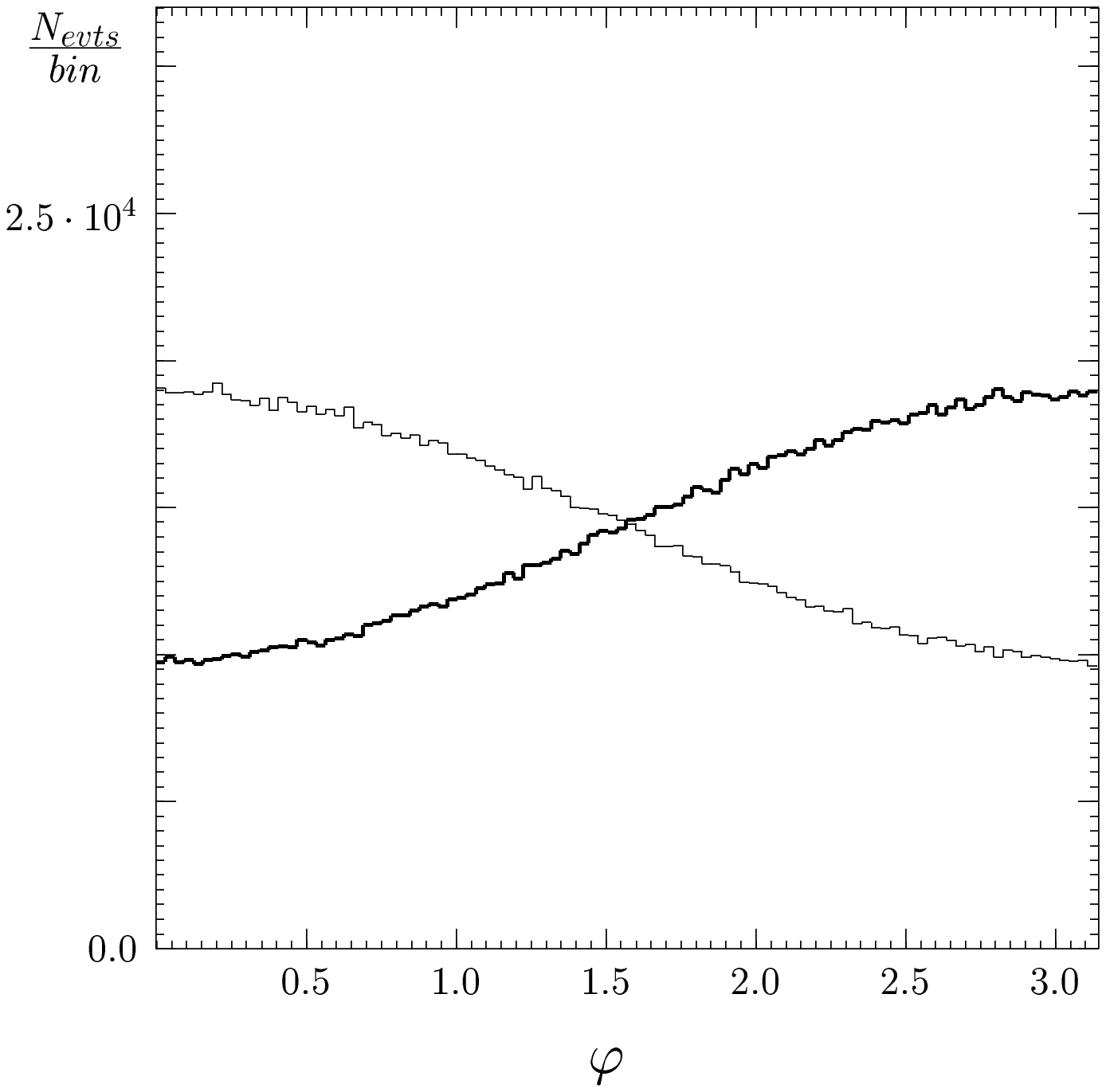,width=120mm,height=140mm}}}
\end{picture}  
\caption  
{\it he $\rho^+ \rho^-$ decay products' acoplanarity distribution
angle, $\varphi$,   in the rest frame of the $\rho^+ \rho^-$
pair,
with cuts depending on the position of the pion with respect to the
$\tau$ cone.  The selection is as in Fig.~\ref{impact2}, but in the
case  of single-solution events (Fig.~\ref{figure2}  case A), the
result from the old method is ignored.  The thick line corresponds to
a scalar Higgs boson,  the thin line to a pseudoscalar one.}
\label{impact3}  
\end{figure}  

\section{Detection parameters}     

To test the feasibility of the measurement, some assumptions  about
the detector  effects have to be made. We include, as the most
critical for our discussion, effects due to    inaccuracies in the
measurements of the   $\pi^\pm,\pi^0$ momenta and $\tau^\pm$ leptons
impact parameters.    We assume Gaussian spreads of the `measured'
quantities with    respect to the generated ones, and we use the
following algorithm to reconstruct the energies of $\pi$'s in their
respective $\tau^\pm$ rest frames.
  
\begin{enumerate}  
\item   
{\bf Charged-pion momentum:}   We assume a 0.1\% spread on its energy
 and direction.
\item   
{\bf Neutral-pion momentum:}   We assume an energy spread of $ 5 \%
\over \sqrt{E [{\rm GeV}]}$. For the $\theta$ and $\phi$ angular    spread
we assume  $ {1 \over 3}  {2 \pi \over 1800}$.  (These neutral pion
resolutions can be    achieved with a 15\% energy error and a
2$\pi$/1800 direction error in the gammas   resulting from the $\pi^0$
decays.)
These resolutions have been approximately verified with a {\tt
SIMDET}~\cite{Pohl:2002vk}, a parametric Monte Carlo program for a
TESLA detector~\cite{Behnke:2001qq}.

\item {\bf The reconstructed Higgs boson rest frame:} We assume a
spread of $2$ GeV with respect to the transverse momentum of the
Higgs boson,  and  $5$ GeV (because of the  beamstrahlung effect)
for the longitudinal component.
\item {\bf The impact parameter: }\\
The angular resolution of the vector pointing from $PP$ to $PCA$ (see
Fig.~\ref{figure1}) has been simulated for the TESLA-like detector. The
simulation is based on the anticipated performance of a 5-layer CCD
vertex detector as described in~\cite{Behnke:2001qq}. For
Higgsstrahlung events with $H\to\tau^+\tau^-$ and $\tau^\pm \to
\rho^\pm\bar{\nu}_\tau(\nu_{\tau})$ at $m_H = $ $120$  GeV and
$\sqrt{s} = $ $350$ GeV, the angular resolution has been found to be
approximately 25$^\circ$. We use this resolution in our simulations.

\end{enumerate}  
\section{Acoplanarity of the $\rho^{+}$ and $\rho^{-}$ decay products with 
the help of the impact parameter }

 For all figures in our paper we will use the same angle $\varphi$
built from smeared four-momenta  of the $\rho^{\pm}$ decay products.
The usefulness of this  acoplanarity distribution manifests itself
only after  selection cuts are applied.

In Fig.~\ref{aco}, our $\rho^+ \rho^-$ decay products acoplanarity
distribution    angle $\varphi$,  defined  in the rest frame of the
reconstructed $\rho^+ \rho^-$ pair is shown. Unobservable
generator-level   $\tau^{\pm}$ rest frames are used for the calculation of
selection cuts  (Eq.~(\ref{selection})).  The two plots represent events
selected by the differences of $\pi^\pm$ $\pi^0$ energies defined in
these respective $\tau^\pm$ rest frames. On the left plot, the  signs
are required to be the same $y_1 y_2 > 0$, whereas on the right one,
events are taken with $y_1 y_2 < 0$. This figure quantifies the size
of the  parity effect in an idealized condition, which we will attempt
to approach with realistic ones.  The size of the effect was
substantially diminished when the detector set-up was included  for
$\tau^\pm$ rest frames reconstruction as well, see Fig.~\ref{acosm2},
following exactly the method of  Ref.~\cite{Bower:2002zx}.  Let us use
this result later, as a reference point  to quantify the size of our
improvements.

It turns out that a slight improvement is visible, when the impact
parameter is taken for the $\tau^\pm$ rest frames determination, as
explained in section 4. In fact, the maximum relative difference
between the scalar and pseudoscalar cases is larger by $\sim 12\%$
(compare  Figs.~\ref{acosm2} and \ref{impact1}) if  a Gaussian spread
of the $\tau$  impact parameter $\sigma_{IP}=25^\circ $  is used.   We
have also checked that the precision of the $\tau$ lepton impact
parameter measurement is not critical in this case. The twofold
ambiguities as explained in Fig.~\ref{figure2}  are the main reason for
the sensitivity to be much lower than in Fig.~\ref{aco}.
 
At the cost of introducing cuts and thus reducing the number of
accepted events, we can go one step further in the improvement of the
method. We first start by requiring that the signs of the energy
differences $y_1$ and $y_2$  be the same with both methods  (with and
without $\tau$ lepton impact parameter).  The result is depicted
in Fig.~\ref{impact2}. Only $\sim 52\%$ of events are accepted, but
the parity effect grows by $\sim 107 \%$ in comparison to
Fig.~\ref{acosm2}.

The events can also be treated differently, according to the number of
solutions; see Fig.~\ref{figure2}. If the  same cut is imposed only for  $\tau^\pm$ leptons
with double solutions, while for single-solution
events the  $\tau^\pm$ rest frames obtained with the help of the impact
parameter are used alone, the number of accepted events turns out to
be  $\sim 72\%$, but the parity effect is larger by only $\sim 67\%$
than in Fig.~\ref{acosm2}.  This result is depicted in
Fig.~\ref{impact3}.

Other cuts are also envisageable, but the two proposed here are  easy to
implement and lead to an acceptable loss of statistics, together with a
sizeable increase of parity effect so that there is not  much room for
further improvements.

\section{Summary}
We have found that the measurement of the impact parameter can be
helpful in the determination of the Higgs boson parity at a future
linear collider, using the $H/A\to\tau^{+}\tau^{-}$;
$\tau^{\pm}\to\rho^{\pm}\bar{\nu}_{\tau}(\nu_{\tau})$ cascade decay.
Thanks to the improved  method the effect of parity is enhanced by $\sim
107\%$ at the cost of losing only $\sim 48\%$  events, which
represents a gain of a factor of $\sim 1.5$ in sensitivity with
respect to the analysis without impact parameter. This
correspond to  $\sim$ 4.5$\sigma$ separation  between scalar and
pseudoscalar for the $120$ GeV Standard Model Higgs  and  other
assumptions similar  to the one taken in Ref.~\cite{Bower:2002zx}.

We have also checked, that reasonable variations in the assumptions of
detection  uncertainties in measurement of charged and neutral $\pi$'s
as well as of the $\tau^+$ and $\tau^-$ impact parameters, hardly change
the sensitivity of our method at all. This is because  the
dominant uncertainty is saturated by the twofold ambiguity as
explained in Fig. 2. We are afraid that a reconstruction of the Higgs
boson  momentum to  precision better than the $\tau$ mass would be
needed to resolve the ambiguity.  This seems  unrealistic for several reasons,
{ e.g.} the beamstrahlung effect.

Probably, a fit to the  multidimensional distribution might be more
efficient than the simple cuts  applied in this work. We believe,
however, that such studies should be done  only when the detector
properties have been understood  better. Finally, we expect that
further improvements, { e.g.}  involving the $\tau^\pm\to
a_{1}^\pm\nu_{\tau}$ decay mode, may turn out to be  helpful for the
measurement of the Higgs boson parity.
 
\subsection*{Acknowledgements}

This work is partly supported 
by the Polish State Committee for Scientific Research  
(KBN) grants Nos  5P03B09320, 2P03B00122 as well as  the
BMBF (WTZ) project number POL~01/103 and by the DESY $\&$ INP 
Cracow collaboration grant TESLA.

One of the authors (M.W.) is grateful for  the warm hospitality
extended to her by the FLC group at DESY in Hamburg.
She also would like  to thank the
``Marie Curie Programme'' of the European Commission  for a fellowship.

Useful discussions  with J. H. K\"uhn
are also acknowledged.
\providecommand{\href}[2]{#2}

\end{document}